\documentstyle[11pt]{article}
\begin{document}
\input{psfig}
\begin{titlepage}
\title{\bf Cosmic Strings in Low Mass Higgs Cosmology}

\author{Warren B. Perkins$^{1}$\footnote{w.perkins@swansea.ac.uk} and
Anne-Christine Davis$^{2}$\footnote{A.C.Davis@damtp.cam.ac.uk} \\ \\
\em ~$^1$Department of Physics, University of Wales Swansea,\\
\em  Singleton Park, Swansea, SA2 8PP, Wales. \\ \\
\em ~$^2$Department of Applied Mathematics and Theoretical Physics,\\
\em University of Cambridge, Cambridge, CB3 9EW, UK.} 

\maketitle

\begin{abstract}
A class of grand unified theories with symmetry breaking scale of order
$10^{16} GeV$ have a Higgs particle with mass in the $TeV$ scale. The cosmology
of such theories is very different from usual. We study the 
cosmic strings obtained in such theories. These strings are much 
fatter than usual and their mass per unit length is reduced, resulting in a 
significant reduction in their cosmological effects. We also study the
temperature evolution of such models. 
\end{abstract}

\vspace{-5.2truein}
\begin{flushright} DAMTP/98-1 \\ SWAT/174 \end{flushright}
\end{titlepage}
\newpage

\section{Introduction}
In recent years there has been much interest in unified fields theories,
in particular, those involving supersymmetry. In a wide class of supersymmetric
grand unified theories some of the phase transitions occur in the $TeV$
range, despite the gauge symmetry breaking scale being of order $10^{16} GeV$.
This occurs in a rather generic class of supersymmetric theories in which
the scalar field potential has `flat directions'. Such theories can arise
in superstring theories \cite {shafi1}. The associated scalar
fields also have mass in the $TeV$ range, usually arising from very small
Higgs self-couplings. However, the gauge bosons have mass of order the
grand unified scale. 

Supersymmetry is looking increasingly likely to be involved in some 
underlying theory that unifies the interactions, whether it is a superstring
theory or a supersymmetric grand unified theory. In such theories, the
so-called flat directions are also a common feature, arising in a class
of theories where the gauge symmetry is broken with a so-called F-term.
Such models could solve the cosmological moduli problem.
Whilst the physical content of such theories is understood, it is only
recently that the cosmology has started to be investigated \cite {Lyth Stewart,
BCLPS,Freeze}. 

In a very innovative paper \cite {Lyth Stewart} Lyth and Stewart considered
the cosmology of models with low mass Higgs particles. They showed that the
evolution of the universe was radically different from usual. In 
addition to the usual inflationary period their model had a period of
thermal inflation followed by a `cold big bang' at $TeV$ scales. This was 
succeeded by Higgs particle decay and a `hot big bang' at around $10 MeV$
just prior to nucleosynthesis.  

In this paper we investigate the cosmology of such theories, considerably
extending previous work. We consider topological defects formed in such
theories, in particular cosmic strings. In these theories the string profiles
are considerably modified by the relatively small Higgs mass. We display the
string profiles in section 2 for both the Higgs and gauge fields and
calculate the string width and mass per unit length. We show that there is
a modification factor depending on the logarithm of the ratio of the scalar 
and gauge particle masses. This modification  results in the strings being
`fat', with decreased mass per unit length. In section 3 we discuss
string dynamics, showing that they are produced after the period of friction
domination, but otherwise evolve as normal strings.
The resulting cosmology of such strings is also considered. We show that, 
if the gauge symmetry breaking scale is around $10^{16} GeV$,
then the strings are less cosmologically significant than usual. We also 
discuss microphysical affects of such strings. 
In section 4 we comment on the temperature-time relation in our model and 
compare it with the temperature evolution discussed previously.
By considering the decay of the Higgs particles in a half-life model, we
show that the dramatic results of \cite {Lyth Stewart} are modified 
considerably.

We summarize our results in section 5.

\section{The Structure of Low Higgs Mass Strings}

In order to consider cosmic strings \cite {HKVS} in grand unified theories
with low mass Higgs particles we can model the 
relevant features of the Higgs potential after supersymmetry breaking
by the usual Mexican hat with a very small Higgs self coupling.
We thus consider an effective Abelian Higgs model with Lagrangian,
$$
{\cal L}=(D^\mu\phi)^\dagger D_\mu\phi -{1\over 4} F_{\mu\nu}F^{\mu\nu}
-\lambda(\phi^\dagger\phi-\eta^2/2)^2.
$$
It should be emphasized that this potential is only appropriate for
tree level calculations.  Loop corrections within the Abelian Higgs
model would usually necessitate a degree of fine tuning to protect 
the very small Higgs self coupling we are considering. However, in the
context of the underlying supersymmetric theory this fine tuning is not
a problem. We can consistently work with the Abelian Higgs model at the
tree level with whatever couplings we choose.
 
We construct a Nielsen-Olesen vortex solution in the usual manner by setting
$$
\phi=\phi(r){\rm e}^{i\theta}
$$ 
and considering only $A_\theta$ non-vanishing. The energy per unit length of 
string is then given by
$$
\epsilon=2\pi\int_0^\infty rdr\bigl[ \phi^2_{,r} 
+\phi^2({1\over r}-eA_\theta)^2 +{1\over 2}(A_{\theta,r}+A_\theta/r)^2
+\lambda(\phi^2-\eta^2/2)^2\bigr]
$$
It is convenient to introduce the scaled variables:
$$
\phi={\eta\over\sqrt{2}}\tilde\phi, \quad A_\theta= \eta\tilde A,
\quad r={x\over \eta e}.
$$
In terms of these variables the energy per unit length reduces to,
$$
\epsilon=\pi\eta^2\int_0^\infty xdx\bigl[ \tilde\phi^2_{,x} 
+\tilde\phi^2({1\over x}-\tilde A)^2 +(\tilde A_{,x}+\tilde A/x)^2
+B(\tilde\phi^2-1)^2\bigr].
$$
where $B=\lambda/2e$

In order to obtain a model for the string we take the following 
forms for the string profiles:
$$
\tilde\phi=\cases{\beta x^\gamma & $0\leq x\leq \beta^{-1/\gamma}$ \cr
                   1              & $ x>\beta^{-1/\gamma}$\cr}
$$
$$
\tilde A=\cases{ {x \over X^2} & $x\leq X$\cr
                  {1\over x}   & $x>X$\cr}
$$
These approximations allow the gauge and scalar cores to have different 
radii and allow the scalar field to either delay its approach to its 
asymptotic form ( large $\gamma$) or move to it more rapidly (small 
$\gamma$).

With these forms for the profile functions the energy per unit length 
becomes,
$$
\epsilon=\pi\eta^2\biggl[{\gamma\over 2} +\beta^2X^{2\gamma}
\bigl({1\over 2\gamma} -{1\over \gamma+1}+{1\over 2\gamma+4}\bigr)
+{2\over X^2} +B\beta^{-2/\gamma}\bigl({1\over 2} 
-{1\over \gamma+1}+{1\over 4\gamma+2}\bigr)\biggr].
$$

We now wish to vary the parameters so as to minimize the string energy. We first
gain a rough idea of how the parameters vary with B and then proceed to a
more careful minimization.

For small values of B, that is small Higgs self coupling,
 we expect the energy to be small. This requires $\gamma$
to be small, X to be
large and $\beta^2/\gamma$ to be small  in order to suppress the individual terms 
in the energy. Assuming that each term in the energy decreases as the inverse of 
some scale $l$ as B becomes small, we have
\def\e{\rm e}
$$
\gamma\sim {1\over l} , \quad X\sim \sqrt{l} ,\quad \beta\sim {1\over l}
$$
and
$$
B\sim l^{-1+2l}
$$
As an approximation we consider 
$$
l= {\log{(1/B)}\over 2\log{\log{(1/B)}}}.
$$ 
With this value of $l$, $B\beta^{-2/\gamma}$ indeed decreases more rapidly than $1/l$,
thus our  model for the
string profile functions gives an energy per unit length that
decreases at least as rapidly as
$$
{2\log\log(1/B)\over \log(1/B)}.
$$

We need to be slightly more careful about the minimization in order to determine how the
core sizes scale with B. We introduce some shorthand,
$$
Q={1\over 2\gamma} -{1\over \gamma+1}+{1\over 2\gamma+4}, \quad
P={1\over 2}-{1\over \gamma+1}+{1\over 4\gamma+2},
$$
then vary the energy with respect to each of the parameters. 
Varying first with respect to X we find 
$$ 
X=\biggl({2\over \beta^2\gamma Q}\biggr)^{1\over 2+2\gamma}.
$$
Eliminating X and varying with respect to $\beta$, we find that the value of $\beta$ is given by
$$
\beta^2=\Biggl[ {BP\over 2}\bigl({Q\gamma\over 2}\bigr)^{-{1\over 1+\gamma}}\Biggr]
^{\gamma(1+\gamma)\over 1+2\gamma}.
$$
We saw above that $\gamma$ is small for small values of B, so we expand our expression for
the energy about $\gamma=0$ and keep only the leading terms; 
$$
\epsilon\simeq\pi\eta^2\biggl[
{\gamma\over 2} 
+{B^\gamma \over 2\gamma}
\biggr].
$$ 
Minimizing with respect to $\gamma$
yields the constraint,

$$
{K^2\over \log^2{B}} +(K-1)\e^K=0,
$$
where $K=\gamma\log{B}$.
There is a positive solution, $K=1+O(1/\log^2{B})$, but this gives an unphysical
negative value to $\gamma$.  For large $\log^2{B}$, the negative solution is
$$
K\simeq -\log(\log^2{B}).
$$
Finally we have
$$
\gamma\simeq {\log(\log^2{(1/B)})\over \log{(1/B)}} ={2\log\log{(1/B)}\over \log{(1/B)}}.
$$
While this coincides with the form for $\gamma$ we obtained by the naive argument,
the forms of $X$ and $\beta$ are different.
Working back through the constraints we find
$$
\beta \simeq B^{\gamma\over 2}={1\over \log(1/B)},\quad
X\simeq 2 \log(1/B).
$$
Fig.1 shows a comparison of the actual Higgs and gauge field profiles 
 of a low Higgs mass string with the model profiles
for $B=10^{-4}$.

Thus, the mass per unit length for these type of cosmic strings is
$$
\mu={\eta^2\over\log{(B^{-1})}}
$$
Hence, for symmetry breaking scale $\eta$ the mass per unit length is
reduced by the logarithmic factor. A similar result was obtained numerically
in ref \cite {BCLPS}, though the full profile functions were not obtained.
For small $\lambda$ this reduction
factor can be an order of magnitude, hence affecting the cosmological 
predictions in this model. Similarly, the gauge and scalar cores are
vastly different. The scalar core size is set by the the inverse Higgs'
mass, whilst the gauge core is increased by $\log{B^{-1}}$. Hence, in
these low mass Higgs models the string width is considerably fatter than
in the usual case. That these strings are much fatter was recognised in
ref \cite {Freeze}, though the reduction in $\mu$ was not realised.

\begin{figure}
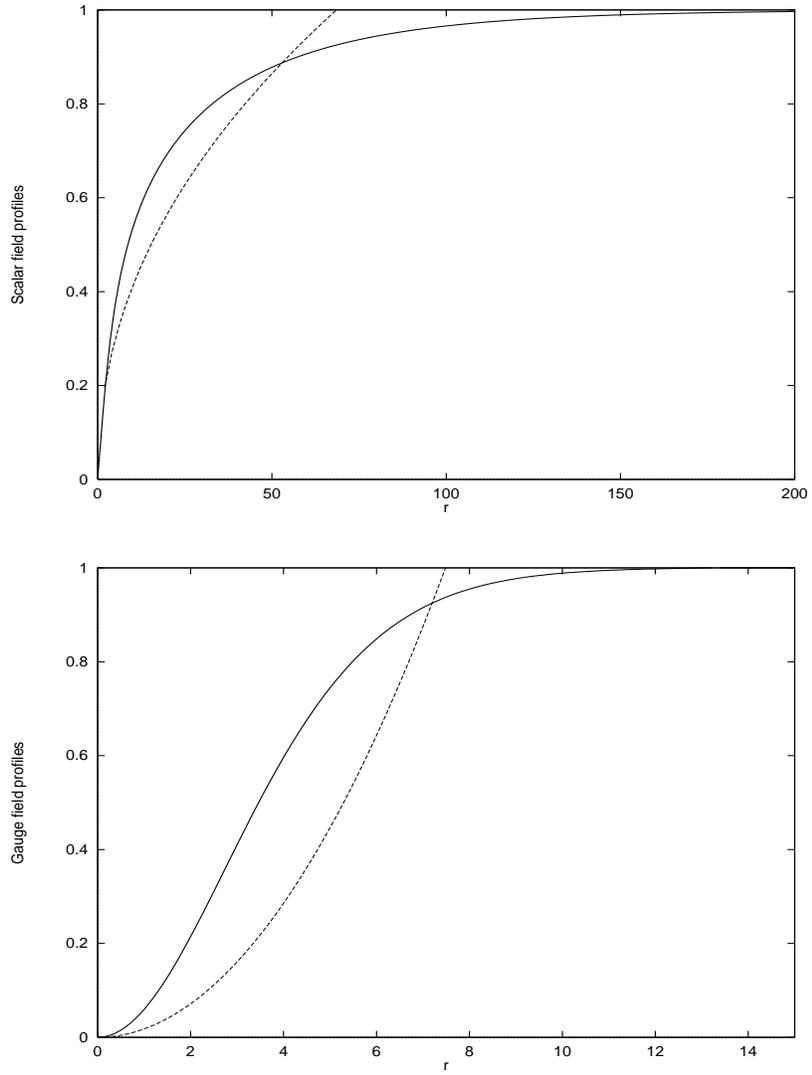

\centerline{\psfig{file=higgs.comp,width=1.5in}}
\centerline{\psfig{file=z.comp,width=1.5in}}
\caption{Comparison of the actual Higgs and gauge field profiles (solid lines) of a low 
Higgs mass string with the model profiles (dashed lines) in this case $B=10^{-4}$. }
\end{figure}

\section{The Evolution and Cosmology of Low Higgs Mass Strings}

In the standard string model there are two distinct periods of string evolution,
depending on the significance of the plasma.
The length scale above which friction is dominant is set by \cite{friction domination}
$$
l_f= {\mu\over \sigma\rho}
$$
where $\mu$ is the mass per unit length of string, $\rho$ is the energy density
of scatterers and $\sigma$ is the scattering cross-section per unit length. While
the mass per unit length of these strings is suppressed by a logarithmic factor,
the scattering cross-section is unchanged. The dominant process is Aharnov-Bohm 
scattering, with the cross-section being determined by the momentum of the scattering
particles, $\sigma\propto T^{-1}$.

In standard string models friction domination ends well before the electroweak
transition. The analysis is similar in this modified picture, with a logarithmic
correction appearing due to the lower mass per unit length. The main difference is
that strings are formed much later, at around the time of the electroweak transition.
Thus strings form well after the period of possible friction domination.
Otherwise their evolution is simply that of normal string with a suitably
reduced mass per unit length.

Whilst the usual Aharonov-Bohm scattering cross-section is the same for
our fat strings as for the normal GUT strings the same is not true for
the inelastic cross-section. For example, the baryon violating catalysis
cross-section will depend on the string radius in realistic GUT models,
unlike the case in toy $U(1)$ models.
This dependence arises from the group generators in the string gauge field
and it also depends on the scattering particle. The full details are shown
in \cite {ad&rj}. Consequently, the catalysis cross-section will be bigger for
fat strings than the corresponding usual GUT strings. This could result 
in some erasing of a primordial baryon asymmetry.

The cosmological effects of cosmic strings are determined by the parameter
$G\mu$. The anisotropy in the microwave background radiation, the density
fluctuations and the gravitational lensing are all proportional to $G\mu$.
With the logarithmic reduction factor, as given in the previous section,
the cosmological effects of these fat strings will be just those of 
ordinary strings formed at a slightly lower energy scale. 
This reduction in the gravitational effects 
of the strings prevents them from playing a major role in structure formation,
unless the unification temperature is slightly raised by a corresponding
amount. Similarly, the power spectrum is likely to be the same shape as for 
the usual GUT scale strings. These strings are likely to
evade constraints arising from microwave background and large scale 
structure measurements. As these strings are appearing in an inflationary 
model, this the lack of string density perturbations is not 
a problem, perturbations arising during the inflationary era can seed large 
scale structure formation. There has been recent interest in mixed models
where the density perturbations are produced by both cosmic strings and
inflationary perturbations. If the cosmic strings are of the fat type
discussed here then their contribution to density perturbations will be
reduced unless unification is raised to compensate for the logarithmic
factor.

The decay of cosmic string loops can result in the production of cosmic rays
and also produce a baryon asymmetry. In \cite {rbad&mh,ad&me} it was shown 
that the decay of GUT scale string loops could account for the resulting 
baryon asymmetry, particularly in models with a low freeze out temperature. The
corresponding case for our fat strings is rather more complicated. This is
because the analysis in \cite {rbad&mh} was performed at the Ginsberg 
temperature
where thermal fluctuations can no longer restore the symmetry. However, since
the fat strings under consideration are produced in first order phase 
transitions, it seems more appropriate to consider the transition temperature
itself. In which case, we have a massive enhancement factor 
coming from the increase in the string width. The results of ref 
\cite{rbad&mh} are enhanced by a factor of $\lambda^{-1}$, in which case
this model can easily account for the observed baryon asymmetry. 

Similarly, there will be a large amount of cosmic rays produced by the
decaying string loops, mainly in the form of $TeV$ Higgs particles. There
may also be cosmic rays produced by the infinite string network \cite {markh}.
Since the particles emitted by the network are going to be mainly $TeV$ scale
Higgs particles and their decay products \cite {shafi}, they will evade the 
bounds of ref \cite {markh}.

However, since the dominant microphysical effects occur  shortly
after string formation, these will be diluted by any subsequent reheating.
If there is sufficient reheating both the baryon asymmetry produced and the
cosmic ray flux will be diluted, and could be diluted below observational
limits. We discuss this in the next section.

\section{The Temperature-Time Relationship For Late Transitions}

The extremely low Higgs mass in these models delays the breaking of the 
GUT symmetry until the temperature is of order the Higgs mass, $\sim 1$ TeV.
Even with the very flat Higgs potential corresponding to such low self couplings,
the vacuum energy density is of order $10^{36} GeV^4$ and is much greater than
the radiation  energy density (of order $10^{12} GeV^4$ at symmetry breaking).
Following symmetry
breaking the vacuum energy density is converted into Higgs particles and the
universe enters a phase of matter domination \cite{Lyth Stewart}. In \cite{Lyth Stewart} it is
argued that this phase lasts for the lifetime of a typical Higgs particle,
of order $10^{23}GeV^{-1}$, then the Higgs particles decay, leading to
 a reheating of the Universe. This reheating produces a standard radiation dominated
epoch that encompasses the final period of nucleosynthesis. 

While this approach provides a reliable estimate of the temperature at which
radiation domination recommences, the temperature evolution during the Higgs
dominated phase is rather different.  It
is more appropriate to think of the Higgs lifetime as a half-life and 
consider a continuous transfer of energy from the matter to radiation
fields.

If $r_d$ is the probability of a Higgs particle decaying per unit time, 
the evolution of the radiation and matter energy densities are  given by
$$
{d \over dt}\rho_r =-4H\rho_r+\rho_m r_d
,\qquad
{d \over dt}\rho_m =-3H\rho_m-\rho_m r_d.
$$
These expressions are valid for both a second order transition \cite{Kolb and Turner}
 and for the first order transition we expect at weak coupling. Bubble
collisions produce a sea of 'soft' quanta \cite{Kolb Riotto and Tkachev}, 
 with energy less than the corresponding instantaneous reheat 
temperature.   Now, even instantaneous conversion of the Higgs' potential energy
to radiation would only give a reheat temperature of around $10^9 GeV$,  much less
than the mass of the particles mediating Higgs decay. Thus  the Higgs decay rate will be
that of static Higgs particles.

We can understand how the energy densities evolve by
considering some approximate solutions.
In a flat, matter dominated  Universe, the evolution of the matter energy density is
given by,
 $$
{d \over dt}\rho_m =-3\sqrt{{8\over 3}\pi G\rho_m}\rho_m-\rho_m r_d.
$$
and we have $\rho_m=\e^{-r_dt}x$ with 
$$
x_f^{-1/2}=x_i^{-1/2}-{3\over r_d}\sqrt{{8\over 3}\pi G}\biggl[\e^{-r_dt_f/2}
-\e^{-r_dt_i/2}\biggr].
$$
To leading order in $r_dt$ this gives
$$
\rho_m=\biggl[x_i^{-1/2}+{3\over 2}\sqrt{{8\over 3}\pi G}(t_f-t_i)\biggr]^{-2}.
$$
Assuming that $x_i$ is very large for $t_i$ close to zero, we have the standard
form,
$$
\rho_m=\biggl[{3\over 2}\sqrt{{8\over 3}\pi G}t\biggr]^{-2}.
$$
As expected, for $r_r t<< 1$ the decay of the Higgs particles has little effect on their
energy density. However, significant energy is transfered into radiation.  
With $\rho_m$ varying as $t^{-2}$, we can solve for the radiation energy density,
$$
\rho_r=at^{-8/3}+ {4\over 15}{r_d \over{8\over 3}\pi G}{1\over t}.
$$
The particular integral varies as $t^{-1}$ and so quickly dominates the complementary
function. Physically, the initial radiation is diluted and redshifted, quickly leaving
only the radiation from the decaying Higgs particles. After a possible  initial 
increase in the temperature, the temperature decreases as $t^{-1/4}$.

Assuming that these small $r_dt$ forms persist until matter-radiation equality,
we have equality at $r_d t =5/3$. Now, the Hubble parameter is given by  $H={2\over 3t}$, 
thus at this
time the Higgs' half-life and the expansion time-scale are comparable and Higgs decay
becomes important in determining $\rho_m$. Following equality the remaining Higgs
particles decay rapidly, but the energy liberated is only of order the radiation
energy density and there is little further reheating. 

We can compare the entropy generated in this approximation with that obtained by
assuming that all of the Higgs' decay at $t=1/r_d$. In both cases we have a phase
of matter domination with $R\propto t^{2/3}$ which ends at $t\sim 1/r_d$. This is 
followed by a phase of radiation domination with an initial radiation energy density,
$$
\rho_r\sim{r^2_d \over{8\over 3}\pi G}.
$$
The entropy generation and temperature at equality are similar in both approximations,
but the temperature evolutions are very different as is the amount of entropy produced
subsequent to any given intermediate temperature.

The forms of $\rho_r$ and $\rho_m$  for  $r_d=10^{-22}GeV$ and the initial 
conditions given above are shown in fig2. It can be seen that there is a small amount
of prompt reheating, but then the universe cools monotonically with matter domination
ending just before $\rho_r$ drops below about $10^{-12} GeV^{-4}$ and 
nucleosynthesis begins.
\begin{figure}
\centerline{\psfig{file=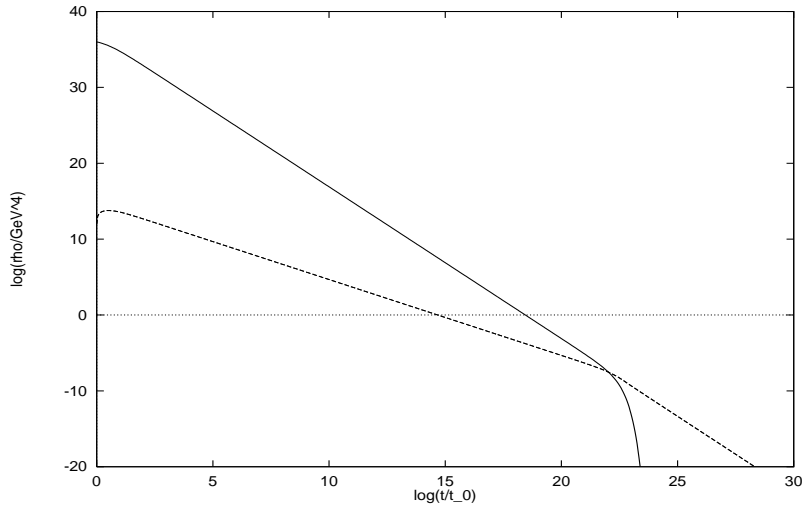,width=1.5in}}
\caption{The Higgs (solid line)
 and radiation energy densities as functions of time}
\end{figure}
\begin{figure}
\centerline{\psfig{file=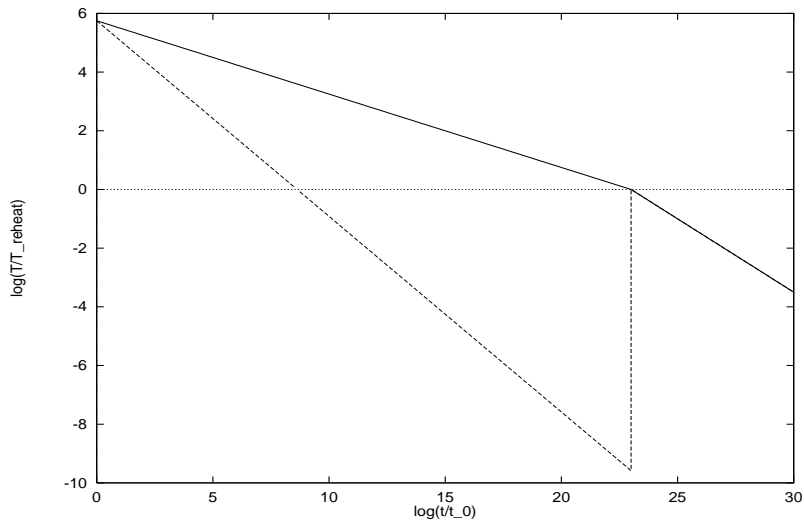,width=1.5in}}
\caption{The temperature as a function of time for continuous Higgs decay 
(solid line) and instantaneous decay.}
\end{figure}

The temperature evolutions for continuous and instantaneous Higgs decay are
compared in fig3. The most significant differences between the two arise
for temperatures above $T_{\rm reheat}$, these temperatures are only 
attained once and occur before much of the entropy generation in the 
instantaneous model. In the continuous decay model these temperatures
are reached later and, particularly in the case of temperatures close to
$T_{\rm reheat}$, there is much less entropy generation at subsequent epochs.

In the instantaneous picture, the temperature increases by a factor
of around $10^{10}$ at reheating. Thus the ratio of the relic density to
the entropy density, $n/s$, for any relic produced at
temperatures above the reheat temperature is reduced by a factor
of around $10^{30}$ at reheating.  In our continuous case, the
scale factor has the usual matter dominated form, $R\propto t^{2/3}$,
while the temperature has the somewhat unusual form, $T\propto t^{-1/4}$.
The relic density to entropy density ratio then has the from,
$$
{n\over s}\propto {R^{-3}\over T^3}\propto t^{-1.25}
$$
For a relic formed at a temperature $T_x$ at time $t_x$, we have a dilution
factor, $d.f.$, of
$$
d.f.=\bigl({ t_{\rm reheat}\over t_x}\bigr)^{1.25}= ({T_x\over T_{\rm reheat}
})^5
$$
For processes occuring at around the TeV scale, the dilution factor is
around $10^{30}$ as in the instantaneous case. However, if the
process occurs at a lower temperature, the dilution factor is reduced. 

In the baryogenesis scenario of \cite {rbad&mh}, the baryon asymmetry was
sufficient to account for that required by nucleosynthesis. Our mechanism
discussed in the previous section enhances this by the factor $\lambda^{-1}$.
However, the dilution factor we have found is considerably greater than this.
Consequently, the resulting baryon asymmetry will be diluted
below that required by nucleosynthesis. A very late baryogenesis mechanism
is needed in this class of models in order to avoid dilution. Similarly,
emitted cosmic rays are likely to be significantly diluted in this class
of models.  

\section{Conclusions}
We have considered the properties of cosmic strings in theories where the
Higgs self-coupling is small. Such theories arise in supersymmetric, grand
unified theories with flat directions. We have shown the resulting 
strings are much fatter than usual, with vastly different gauge and scalar
field core sizes. The gauge core is increased by a logarithmic factor, whilst
the scalar core is the inverse Higgs mass. Similarly, the mass per unit 
length of the string is reduced by the logarithmic suppression factor 
relative to usual cosmic string models.
This alters the cosmology of such strings. The gravitational properties
are reduced by this factor. Consequently, if such strings are to play a
role in large scale structure then the unification temperature needs to be
increased by a corresponding amount. This makes such models less attractive
from a particle physics viewpoint, though they could still arise from 
superstring theories. Similarly, the increase in string width changes their
microphysical properties. It results in an increased baryon catalysis 
cross-section and a vast increase in the number of particles produced by
loop decay. Whilst this latter effect could produce the observed baryon 
asymmetry and result in massive production of cosmic rays from such strings,
it is likely that such observable effects will be diluted by the subsequent
reheating. 

The evolution of the universe is greatly modified in this class of theories,
with the universe undergoing a late reheating. Using a realistic model 
for the decay of the $TeV$ scale Higgs particle, we have shown how the
evolution is modified. There is prompt reheating following the phase 
transition, the universe then undergoes a period of matter domination
as it cools monotonically until radiation domination begins just before
nucleosynthesis. This scenario is rather different from that proposed in
\cite {Lyth Stewart} where all Higgs particles were taken to decay at the 
typical lifetime of order $10^{23}GeV^{-1}$. In particular, relics produced
at temperatures above the reheat temperature suffer significantly less
dilution if the Higgs decay is continuous than they would if the decay was
instantaneous. 

\section*{Acknowledgements}
This work was supported in part by PPARC. This work is in part the result
of a scientific network supported by the European Science Foundation.
WBP would like to thank R.A.W. Gregory for useful discussions.

\end{document}